\documentstyle[prl,floats,aps,epsf]{revtex}
\input epsf

\begin{document}
\twocolumn[\hsize\textwidth\columnwidth\hsize\csname@twocolumnfalse\endcsname

\draft
\title{Metal-Insulator Transition and Spin Degree of Freedom \\
 in Silicon 2D Electron Systems}
\author{T. Okamoto,$^1$ K. Hosoya,$^1$ S. Kawaji,$^1$ A. Yagi,$^2$ A. Yutani$^3$ and Y. Shiraki$^3$}
\address{$^1$Department of Physics, Gakushuin University, Mejiro, Toshima-ku, Tokyo 171-8588, Japan}
\address{$^2$NPC Ltd., Shiobara-cho, Tochigi 329-2811, Japan}
\address{$^3$RCAST, The University of Tokyo, Komaba, Meguro-ku, Tokyo 153-0041, Japan}
\date{June 30, 1999}
\maketitle

\begin{abstract}
Magnetotransport in 2DES's formed in Si-MOSFET's and Si/SiGe quantum wells at low temperatures is reported.
Metallic temperature dependence of resistivity is observed for the {\it n}-Si/SiGe sample even in a parallel magnetic field of 9~T, where the spins of electrons are expected to be polarized completely.
Correlation between the spin polarization and minima in the diagonal resistivity observed by rotating the samples for various total strength of the magnetic field is also investigated.

\end{abstract}

\pacs{71.30.+h, 73.40.Qv, 73.20.Dx}
]

Metallic temperature dependence of resistivity at a zero magnetic field has been observed in two-dimensional electron systems (2DES's) in Si-MOSFET's \cite{KRAV1,KRAV2} and other 2D systems \cite{SIGE1,SIGE2,GAAS1,GAAS2,ALAS,GAAS3} characterized by strong Coulomb interaction between electrons (or holes) \cite{ALAS}.
In experiments on Si-MOSFET's, it was found that a magnetic field applied parallel to the 2D plane can suppress the metallic behavior \cite{KRAV3,PUD1,OKA3}.
This result indicates that the spins of electrons play an important role in the metallic region as well as in the insulating region \cite{OKA1,OKA2}.

The parallel magnetic field $ B _ { \parallel } $ does not couple the orbital motion of electrons within the 2D plane, but it changes the spin polarization of the 2DES.
The spin polarization can be defined as $ p = ( N _ { \uparrow } - N _ { \downarrow } ) / N _ { s } $, where $ N _ { \uparrow } $ and $ N _ { \downarrow } $ are the concentrations of electrons having an up spin and a down spin, respectively, and $ N _ { s } $ is the total electron concentration ($ N _ { s } = N _ { \uparrow } + N _ { \downarrow } $).
$ p $ is expected to increase linearly with the total strength $ B _ { \rm tot } $ ($ = B _ { \parallel } $) of the magnetic field.
We have $ p = B _ { \rm tot } / B _ { c } $ for $ p < 1 $ and $ B _ { c } = 2 \pi \hbar ^ { 2 } N _ { s } / \mu _ { B } g _ { v } g _ { \rm FL } m _ { \rm FL } $ if the system can be considered as Fermi liquid.
Here, $ \mu _ { B } ( = \hbar e / 2 m _ { e } ) $ is the Bohr magneton and the valley degeneracy $ g _ { v } $ is 2 on the (001) surface of silicon.
The strong {\it e-e} interaction is expected to change the effective $g$-factor $ g _ { \rm FL } $ and the effective mass $ m _ { \rm FL } $ from $ g ^ { \ast } = 2.0 $ and $ m ^ { \ast } = 0.19 m _ { e } $ in the non-interacting 2DES in a (001) silicon surface.

In the present work, we use two $n$-type silicon samples.
A Si-MOSFET sample denoted Si-M has a peak electron mobility of $ \mu _{ \rm peak} = 2.4~{ \rm m }^{2} / { \rm V~s} $ at $ N _ { s } = 4 \times 10 ^ { 15 } { \rm m } ^ { -2 } $ and $ T = 0.3~{ \rm K } $.
The estimated SiO$_2$ layer thickness is 98~nm.
A Si/SiGe sample denoted Si-G was grown by combining gas-source MBE and solid-source MBE.
Details of the growth and characterization have been reported elsewhere \cite{YUTANI1,YUTANI2}.
The thickness of the strained silicon channel layer is 20~nm.
It is sandwiched between relaxed Si$_{0.8}$Ge$_{0.2}$ layers and separated from a Sb-$ \delta $-doped layer by a 20~nm spacer.
The electron concentration $ N _ { s } $ can be controlled by varying the substrate bias voltage above 20~K.
It has a high mobility of $ \mu = 66~{ \rm m }^{2} / { \rm V~s} $ at $ N _ { s } = 2.2 \times 10 ^ { 15 } { \rm m } ^ { -2 } $ (at zero substrate bias voltage) and $ T = 0.36~{ \rm K } $.
The samples were mounted on a rotatory thermal stage in a pumped $^3$He refrigerator or in a $^3$He-$^4$He dilution refrigerator together with a GaAs Hall generator and resistance thermometers calibrated in magnetic fields.

In Ref.~\onlinecite{OKA3}, some of the present authors have determined the product of $ g _ { \rm FL } $ and $ m _ { \rm FL } $ in Si-M from the low-temperature Shubnikov-de Haas oscillations in tilted magnetic field based on work of Fang and Styles \cite{FS}.
\begin{figure}
\centerline{
\epsfysize=95mm
\epsfbox{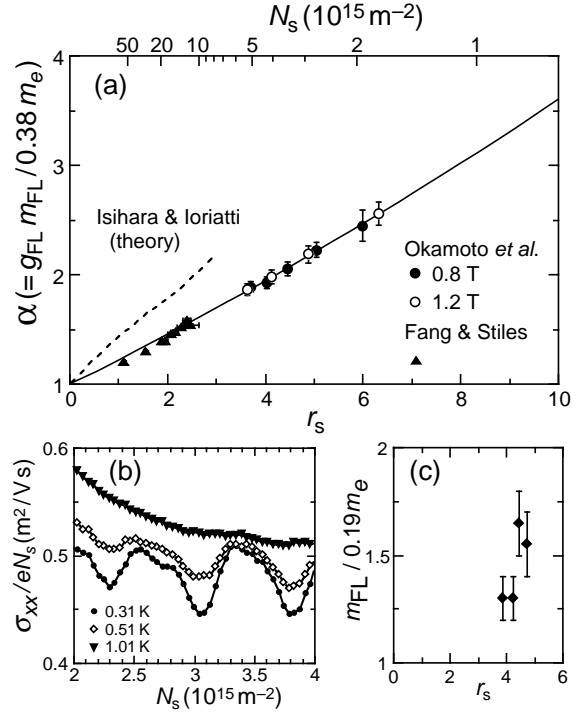}
}
\vspace{3mm}
\caption{ (a) The enhancement factor $ \alpha = ( g _ { \rm FL } / g ^ { \ast } ) / ( m _ { \rm FL } / m ^ { \ast } ) $. (b) Temperature dependence of the Shubnikov-de Haas oscillation at $ B _ { \rm tot } = B _ { \perp } = 0.8~{\rm T} $. (c) The enhancement factor for the effective mass ($ = m _ { \rm FL } / m ^ { \ast } $).
}
\label{FGOS}
\end{figure}
The obtained enhancement factor $ \alpha = g _ { \rm FL } m _ { \rm FL } / 0.38 m _ { e } $ is shown in Fig.~\ref{FGOS}(a) as a function of $ r _ { s } = \pi ^{ 1 /2 } ( e / h )^2 ( m^{\ast} / \kappa \varepsilon_0 ) N_{\rm s} {}^{-1/2} $.
$ r _ { s } $ is a dimensionless parameter indicating the ratio of the Coulomb energy to the Fermi energy.
Here, $ m^{\ast} = 0.19 m _ { e } $ and the relative dielectric constant $ \kappa = 7.7 $ are used.
The dashed line represents calculation based on Ref.~\onlinecite{II}.
The previous experiment could not determine $ \alpha $ for the low-$ N _ { s } $ region, while the metallic behavior was clearly observed in this sample for $ 1 \leq N _ { s } (10 ^ { 15 } {\rm m} ^ { -2 } ) \lesssim 2 $ \cite{OKA3}.
We assume a simple extrapolation function 
\begin{eqnarray}
\alpha - 1 = 0.2212 r _ { s } + 0.003973 r _ { s } {} ^ { 2 } .
\label{Hex}
\end{eqnarray}
represented by the solid line in Fig.~\ref{FGOS}(a).

Fig.~\ref{FGOS}(b) shows typical temperature dependence of the Shubnikov-de Haas oscillation in a perpendicular field of 0.8~T.
The effective mass roughly estimated is plotted in Fig.~\ref{FGOS}(c).
The effective $ g $-factor is expected to be also enhanced by the $ e $-$ e $ interaction since $ \alpha = ( g _ { \rm FL } / g ^ { \ast } ) / ( m _ { \rm FL } / m ^ { \ast } ) $ in Fig.~\ref{FGOS}(a) is larger than $ m _ { \rm FL } / m ^ { \ast } $.

\begin{figure}
\vspace{10mm}
\centerline{
\epsfysize=75mm
\epsfbox{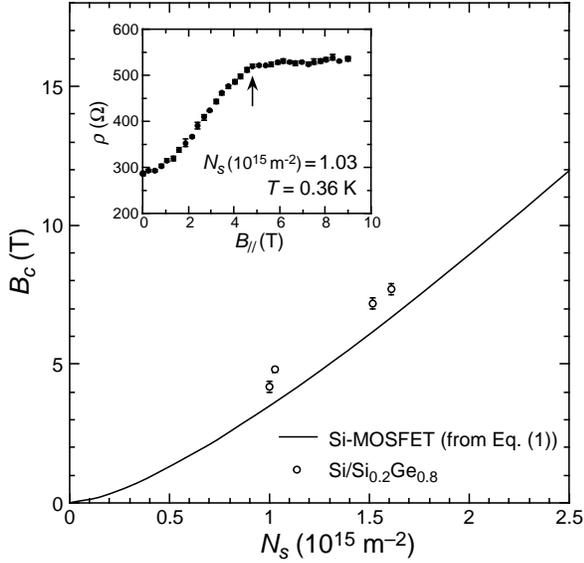}
}
\vspace{3mm}
\caption{Solid line represents $ B _ { c } $ for Si-MOSFET's calculated using~(\protect\ref{Hex})~(see text). Open circles are determined from the kink that appears in the $ \rho $ vs $ B _ { \parallel } $ curve for Si-G as shown in the inset.
}
\label{FGBCR}
\end{figure}

The critical magnetic field $ B _ { c } = 2 \pi \hbar ^ { 2 } N _ { s } / \mu _ { B } g _ { v } g _ { \rm FL } m _ { \rm FL } $ for
the complete spin polarization is calculated using (\ref{Hex}) and shown as the solid line in Fig.~\ref{FGBCR}.
It was found that the low-temperature resistivity of Si-MOSFET's increases with the parallel magnetic field and it saturates at high fields \cite{KRAV3,PUD1,OKA1,OKA2}.
The saturation has been related with the complete spin polarization at $ B _ { c } $ \cite{OKA3}.
Similar magnetoresistance is also observed in Si-G, while the $ B = 0 $ resistivity is much smaller than that reported for Si-MOSFET's.
Typical data are shown in the inset to Fig.~\ref{FGBCR}.
A sharp kink is observed at 4.8~T.
The critical magnetic field in Si-G determined from the kink (open circles) is slightly higher than the solid line for Si-MOSFET's.
We consider that the deviation arises from the difference in the effective dielectric constant in the proximity to the interface.

\begin{figure}
\centerline{
\epsfysize=70mm
\epsfbox{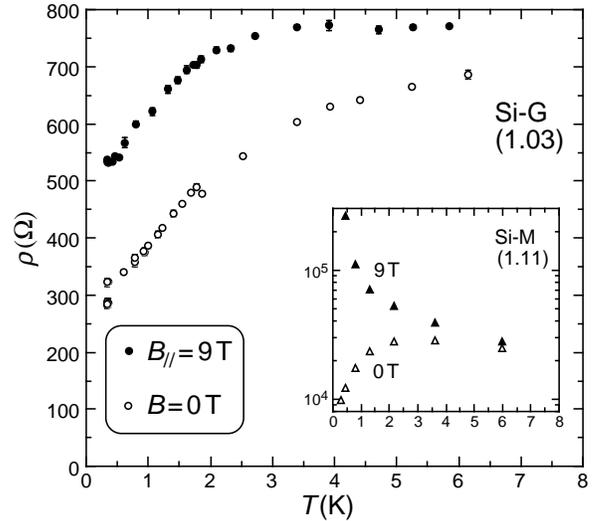}
}
\vspace{3mm}
\caption{Temperature dependence of resistivity of Si-G at $ N _ { s } = 1.03 \times 10 ^ { 15 } {\rm m } ^ { -2 } $ in a zero magnetic field and a parallel magnetic field of 9~T.
The inset shows the data for Si-M at $ N _ { s } = 1.11 \times 10 ^ { 15 } {\rm m } ^ { -2 } $.
}
\label{FGPARA}
\end{figure}

Figure~\ref{FGPARA} shows temperature dependence of the resistivity.
The metallic temperature dependence ($ d \rho / d T > 0 $) in a zero magnetic field is observed in Si-G as well as in Si-M.
The temperature dependence of $ \rho $ in Si-M changes to insulating in a parallel magnetic field (see the inset) as reported on other Si-MOSFET's \cite{KRAV3,PUD1}.
On the other hand, the metallic temperature dependence is observed for Si-G even at $ B _ { \parallel } = 9~{ \rm T } $, where the spins of electrons are expected to be polarized completely.
The result indicates that the metallic behavior can appear even if the spin degree of freedom is limited.

Tilting the 2DES samples from the plane parallel to the magnetic field, one can apply the perpendicular component $ B _ { \perp } $ of the magnetic field independently of the total strength $ B _ { \rm tot } $.
Measurements on $ \rho _ { xx } $ in the low-$ B _ { \perp } $ region for various fixed values of $ B _ { \rm tot } $ using Si-M have been reported in Ref.~\onlinecite{OKA3}.
Unlike ordinary Shubnikov-de Haas oscillations which show $ \rho _ { xx } $ minima at fixed points with $ N _ { \phi } / N _ { s } = { \rm integer } ^ { - 1 } $ ($ N _ { \phi } = e B _ { \perp } / h $), the $ N _ { \phi } / N _ { s } $ values at the $ \rho _ { xx } $ minima depend on the total strength $ B _ { \rm tot } $.

\begin{figure}
\centerline{
\epsfysize=80mm
\epsfbox{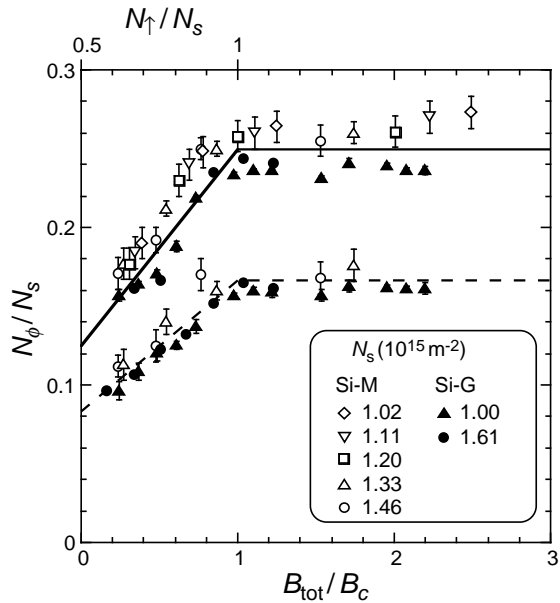}
}
\vspace{3mm}
\caption{Positions of the $ \rho _ { xx } $ minima for different values of $ N _ { s } $ in Si-M (Ref.~\protect\onlinecite{OKA3}) and Si-G are plotted.
Solid line represents $ N _ { \uparrow } / N _ { \phi } = 4 $.
Dashed line represents $ N _ { \uparrow } / N _ { \phi } = 6 $.
}
\label{FGMIN}
\end{figure}

Similar behavior was also observed for Si-G \cite{TOBE}.
In Fig.~\ref{FGMIN}, the positions of the $ \rho _ { xx } $ minima for Si-M and Si-G are shown.
$ B _ { \rm tot } $ is normalized by $ B _ { c } $ presented in Fig.~\ref{FGBCR}.
All the data are distributed along the solid line or the dashed line.
The value of $ N _ { \phi } / N _ { s } $ increases linearly with $ B _ { \rm tot } / B _ { c } $ for $ B _ { \rm tot } / B _ { c } < 1 $, but saturates for $ B _ { \rm tot } / B _ { c } \geq 1 $ where the spin polarization is expected to be completed.

We link this behavior to the concentration $ N _ { \uparrow } $ of electrons having an up spin.
Assuming that a change in $ p $ by $ N _ { \phi } / N _ { s } $ is negligible in the low-$ N _ { \phi } / N _ { s } $ region, we have $ N _ { \uparrow } / N _ { \rm s } = (1 + B _ { \rm tot } / B _ { c } ) / 2 $ for $ B _ { \rm tot } / B _ { c } < 1 $ and $ N _ { \uparrow } / N _ { \rm s } = 1 $ for $ B _ { \rm tot } / B _ { c } \geq 1 $.
The number of ``spin-up'' electron per flux quantum takes constant value of $ N _ { \uparrow } / N _ { \phi } = 4 $ on the solid line or $ N _ { \uparrow } / N _ { \phi } = 6 $ on the dashed line.

The data plotted in Fig.~\ref{FGMIN} have been obtained in the metallic region and the insulating region.
The resistivity at $ B _ { \perp } = 0 $ ranges from $ 10 ^ {2} ~ \Omega $ to $ 10 ^ {6} ~ \Omega $ \cite{OKA3,TOBE}.
An interpretation based on the Shubnikov-de Haas oscillation seems to be hard to apply for the high resistivity region since the Landau level separation is considered to be entirely smeared out by the level broadening \cite{OKA3}.
We believe that the $ \rho _ { xx } $ minima linked with $ N _ { \uparrow } $ is inherent in the strongly correlated electron systems which show the metal-insulator transition.

In summary, we have studied the magnetotransport in 2DES's formed in silicon inversion layers using two samples having quite different electron mobilities.
Positive temperature dependence of the resistivity was observed for a Si/Si$_{0.8}$Ge$_{0.2}$ quantum well even in a parallel magnetic field of 9~T, while the strong parallel magnetic field eliminates the metallic behavior in Si-MOSFET's.
We also investigated the $ \rho _ { xx } $ minima observed in the $ B _ { \perp } $-dependence.
It was related to the concentrations of ``spin-up'' electrons.

This work is supported in part by Grants-in-Aid for Scientific Research from the Ministry of Education, Science, Sports and Culture, Japan, and High-Tech-Research Center in Gakushuin University.

\end{document}